\documentclass[twocolumn,letterpaper,aps,prc,superscriptaddress,nofootinbib,showpacs,floatfix]{revtex4-1}
\usepackage{graphicx}
\usepackage{comment}
\usepackage{setspace}
\usepackage{amsmath}
\usepackage{amssymb}
\usepackage[textwidth=16cm,textheight=22cm]{geometry}
\usepackage{color}
\usepackage{indentfirst}
\usepackage{xspace}
\usepackage{verbatim}
\usepackage{epstopdf}

\usepackage{lineno}

\newcommand{\sqs}{\mbox{$\sqrt{s}$}\xspace}
\newcommand{\ee}{\mbox{$e^{+} e^{-}$}\xspace}
\newcommand{\pp}{\mbox{$pp\,(p\bar{p})$ }\xspace}

\begin{document}

\title{Multiplicity distributions in \ee collisions using Weibull distribution}
 \author{Sadhana Dash} \author{Basanta K. Nandi} 
\author{Priyanka Sett}
\
\affiliation{Indian Institute of Technology Bombay, Mumbai, India}

\email{sett.priya@phy.iitb.ac.in}

\date{\today}  


\begin{abstract}

The two parameter Weibull function is used to describe the charged
particle multiplicity distribution in \ee collisions at the highest
available energy measured by TASSO  and ALEPH
experiments. The Weibull distribution has wide applications in
naturally evolving processes based on fragmentation and sequential branching.
The Weibull model describes the multiplicity distribution very well, as particle
production processes involve QCD parton fragmentation. The effective
energy model of particle production was verified using Weibull
parameters and the same was used to predict the multiplicity
distribution in \ee collisions at future collider energies.

\end{abstract}

\maketitle
\section{Introduction}
\label{intro}
The charged particle multiplicity distribution is one of the most
basic measurements performed in high energy leptonic collision
experiments. This particular measurement provides an insight 
to the multi-particle production mechanism. The formulation of this multi-particle 
production is a complex task. So, one has to rely on model studies
which are based on  quantum chromodynamics as well as a
``soft''  physics which has a significant contribution towards particle 
production. 


Several model studies based on perturbative quantum chromodynamics (pQCD) approach as well as semi-classical approach had been done to understand this complex task~\cite{1,2,3, books}.  Most of the theoretical models which has a pQCD contribution 
based on parton fragmentation and sequential branching has successfully 
explained the measured data from LEP experiments~\cite{ALEPH1, OPAL, DELPHI1}. The particle production 
in such models has an iterative branching process in which the initial quarks 
produced in \ee collisions radiate gluons which inturn branch into cascade 
of partons until the virtualities become negligible to allow further branching~\cite{physlettb282_1992_471}. This is followed by hadronization. 
  
 The multiplicity distribution follows a Poisson distribution 
 if the final state of particles are produced independently. 
 In experiments with higher center of mass energies and different rapidity 
ranges, it was seen that the shape of the multiplicity 
distribution  deviates from the Poissonian shape. The Negative Binomial Distribution
(NBD)~\cite{NBD, NBD1}, which has two parameters, namely, $k$ (measures
deviation from Poisson distribution) and $\langle n \rangle$ (average
number of particles),  successfully described  the particle
multiplicity both in \pp~\cite{NBD}  and \ee~\cite{eeNBD} 
collision systems. However, the NBD failed to provide a good description of 
the multiplicity distributions at higher energies and this deviation 
was also observed in LEP experiments~\cite{ALEPH1, OPAL, DELPHI1, physlettb282_1992_471, spaper}. 
In view of this several other distributions such as Modified Negative 
Binomial Distribution (MNBD)~\cite{MNBD}  and 
Log-Normal~\cite{LND} distributions emerged and successfully described the 
data.   A nice description of multiplicity distributions and various approaches can be found in Ref.~\cite{mp1, mp2, mp3, KReygers}. 

Since the hadron multiplicities in \ee collisions can be understood as an 
outcome of a broad class of branching processes, Weibull distribution~\cite{weibull, weibull1, brownpaper} 
can be used to describe the distribution of produced charged
particles. 
This distribution has successfully described the multiplicity distribution
in \pp collisions for a broad range of center of mass energies and
pseudorapidity intervals \cite{weibull1}. As \ee collisions are more
fundamental and less complex than the \pp collisions and a sequential branching 
is one of the major process for particle production, it will be
interesting to see whether Weibull function describes the 
measured multiplicity distributions in leptonic collisions for broad
range of energies.

 In the present paper, Weibull function is used to describe the
 multiplicity distribution in \ee collisions measured by
 TASSO~\cite{TASSO} and ALEPH~\cite{ALEPH} experiments at PETRA  and
 LEP energies, respectively.  
Although there are several statistical models which describe the 
charged particle distribution in \ee collisions, the idea of this 
present paper is to test the applicability of Weibull distribution 
in such collision systems and interpret the mechanism of particle production 
in terms of its parameters.

\section{Weibull distribution}
\label{formulation}

Many evolving systems are found to show a skewed behaviour.   These kind of 
systems are well described by power law as well as log-normal distribution and Weibull distribution. 
 A detailed review of its applications are given in Ref.~\cite{weibull1, brownpaper}. 

 The Weibull distribution for a random variable `$n$' is expressed as, 
 \begin{equation}
 \label{eq1}
 P(n, \lambda, k) = \frac{k}{\lambda} \, \left(\frac{n}{\lambda}\right)^{k-1} \, {e^{-(n/\lambda)}}^{k}
 \end{equation}
 Here $k$ is known as the shape parameter and $\lambda$ is known as the 
 scale parameter of the distribution. The mean of the distribution is expressed as; 
 \begin{equation}
 \label{eq2}
 \langle n \rangle = \lambda \, \Gamma\left(1 + \frac{1}{k}\right)
 \end{equation}
 
Weibull distribution can be used to describe  the multiplicity distribution obtained from 
the hadronic or leptonic collisions at high energy, as the underlying mechanism is based on initial parton fragmentation and successive branching.

\section{Results and Discussions}
\label{results}

The multiplicity distributions measured in the \ee collisions 
by TASSO~\cite{TASSO} and ALEPH~\cite{ALEPH} experiments are
fitted with Weibull function. 
The parameters of Weibull distribution are studied   for two extreme
rapidity intervals  ($|y| <$ 2.0 and $|y| <$ 0.5) at various \sqs (14 GeV, 22 GeV, 34.8 GeV, 43.6 GeV and 91.2 GeV).

\begin{figure}
\includegraphics[width=0.48\textwidth]{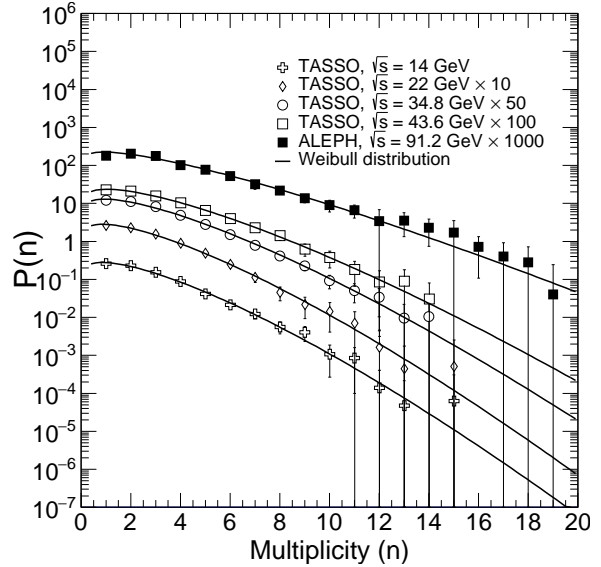}
\caption{\label{f1} Multiplicity distribution for $|$y$|$ $<$ 0.5 measured at \sqs = 14,
  22, 34.8 and 43.6 GeV  by TASSO~\cite{TASSO} collaboration and \sqs = 91 GeV by
  ALEPH~\cite{ALEPH} collaboration. The solid line represents the Weibull fit to the data points. The data points for a given energy are appropriately scaled for better visibility. }
\end{figure}

\begin{figure}
\includegraphics[width=0.48\textwidth]{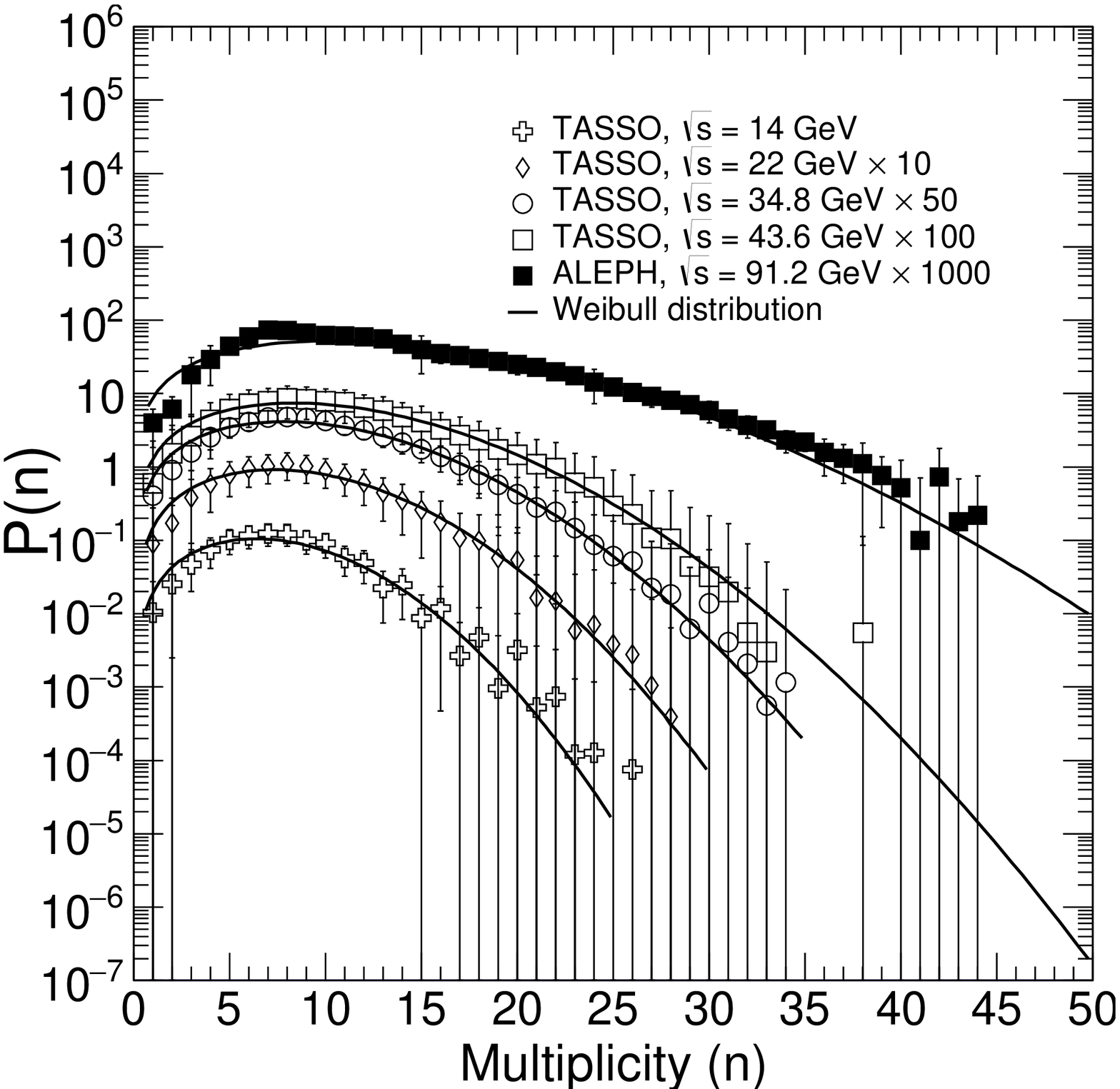}
\caption{\label{f2} Multiplicity distribution for $|$y$|$ $<$ 2 measured at \sqs = 14,
  22, 34.8 and 43.6 GeV  by TASSO~\cite{TASSO} collaboration and \sqs = 91.2 GeV by
  ALEPH~\cite{ALEPH} collaboration. The solid line represents the Weibull fit to the data points. The data points for a given energy are appropriately scaled for better visibility. }
\end{figure}

\begin{table*}
\caption{Weibull parameters obtained from multiplicity distribution for
  $|y|<$ 0.5 measured at   \sqs = 14,  22, 34.8 and 43.6 GeV  by 
TASSO~\cite{TASSO} collaboration and at \sqs = 91 GeV by  ALEPH~\cite{ALEPH} collaboration.}
\label{t1}
\begin{tabular}{ccccc}
\hline
\hline
\sqs &$k$&$\lambda$&$\langle{n}\rangle$&$\chi^{2}$/NDF\\
(GeV)&&&&\\
\hline
14.00 \, & 1.33 $\pm$ 0.06 \, &  2.42 $\pm$ 0.06 \, &  2.22 $\pm$ 0.05  & 0.38 \\
 22.00 \, & 1.36 $\pm$ 0.07 \, &  2.47 $\pm$ 0.06 \, &  2.26 $\pm$ 0.06  & 0.13 \\
 34.80 \, & 1.34 $\pm$ 0.03 \, &  2.66 $\pm$ 0.04 \, &  2.45 $\pm$ 0.03  & 0.70 \\
 43.60 \, & 1.35 $\pm$ 0.06 \, &  2.95 $\pm$ 0.11 \, &  2.71 $\pm$ 0.08  & 0.07 \\
 91.2 \, & 1.25 $\pm$ 0.07 \,  &  3.31 $\pm$ 0.17 \, &  3.08 $\pm$ 0.11  & 0.39 \\
\hline
\hline
\end{tabular}
\end{table*}

Figure~\ref{f1} and Figure~\ref{f2} show the charged particle multiplicity distribution for 
$|y| <$ 0.5  and $|y| <$ 2.0, respectively,  fitted with Weibull
distribution. The distributions 
corresponding to different \sqs are shown by different markers and are scaled by a suitable factor (see Figure.~\ref{f1} and Figure.~\ref{f2}) for visual
clarity. The solid lines are the Weibull fits to the data points. It
can be observed from  both the figures that the 
Weibull distribution describes the data 
very nicely. The fitting is performed using the $\chi^2$ minimization 
method. The Weibull parameters $k$ and $\lambda$ 
along with the $\chi^{2}$/NDF
and $\langle{n}\rangle$  are listed in Table~\ref{t1} and 
Table~\ref{t2} for $|y| < $ 0.5 and $|y| < $ 2.0, respectively.

\begin{table*}
\caption{Weibull parameters obtained from multiplicity distribution for
  $|y|<$ 2 measured at   \sqs = 14,  22, 34.8 and 43.6 GeV  by 
TASSO~\cite{TASSO} collaboration and at \sqs = 91 GeV by  ALEPH~\cite{ALEPH} collaboration.}
\label{t2}
\begin{tabular}{ccccc}
\hline
\hline
\sqs &$k$&$\lambda$&$\langle{n}\rangle$&$\chi^{2}$/NDF\\
 (GeV)&&&&\\
\hline
14.00 \, & 2.20 $\pm$ 0.08 \, &  8.39 $\pm$ 0.37  \,&  7.43 $\pm$ 0.33  & 0.19 \\
 22.00 \, & 2.18 $\pm$ 0.07 \, &  9.70 $\pm$ 0.52 \, &  8.59 $\pm$ 0.46  & 0.07 \\
 34.80 \, & 2.10 $\pm$ 0.04 \, &  10.62 $\pm$ 0.33 \, &  9.41 $\pm$ 0.29  & 0.14 \\
 43.60 \, & 2.05 $\pm$ 0.18 \, &  11.59 $\pm$ 0.64 \, &  10.27 $\pm$ 0.56  & 0.02 \\
 91.00 \, & 1.96 $\pm$ 0.04 \, &  14.97 $\pm$ 0.22  \,&  13.74 $\pm$ 0.19  & 1.51 \\
 
\hline
\hline
\end{tabular}
\end{table*}

\begin{figure}
\includegraphics[width=0.48\textwidth]{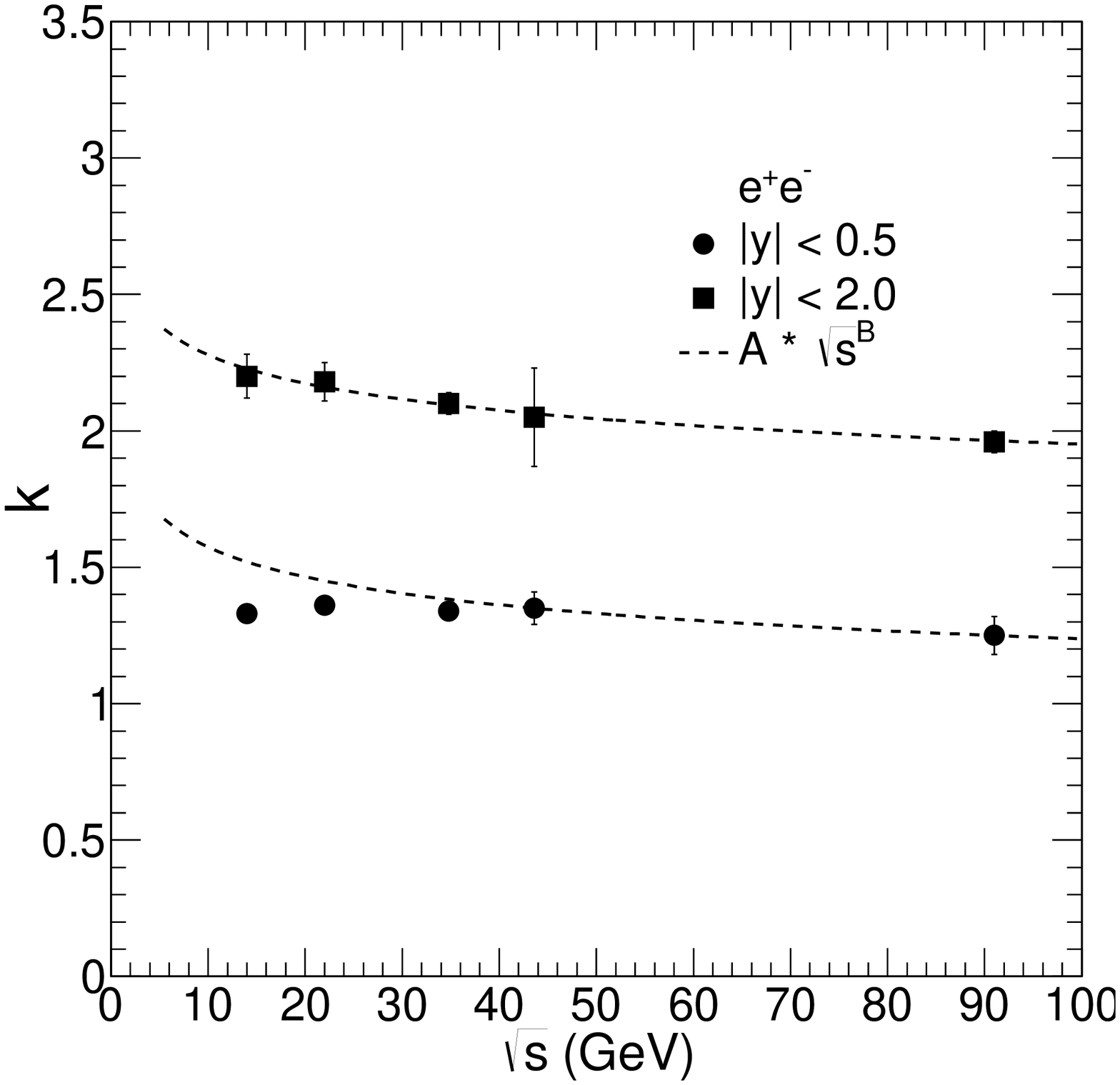}
\caption{The variation of parameter $k$ as  a function 
        	of $\sqrt{s}$ and rapidity. The parameter $k$ is
        parameterized with a power law of the form 
        A$\times$ $\sqrt{s}^{B}$, shown by the dashed line. 
        The parameter $k$ is extracted from the \ee 
        	data~\cite{TASSO, ALEPH}.}
\label{fig:k}
\end{figure} 
 	
\begin{figure}
\includegraphics[width=0.48\textwidth]{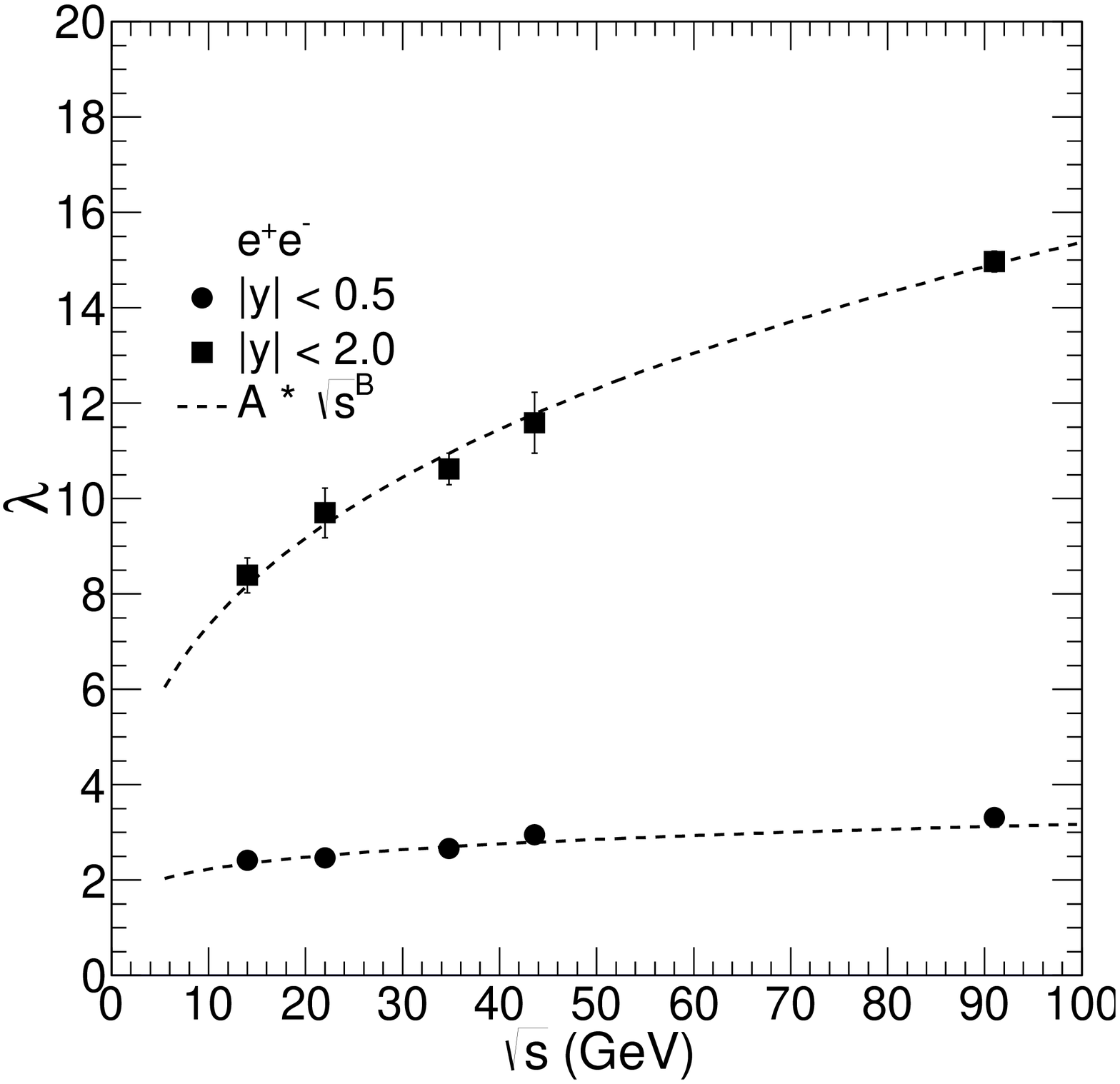}
\caption{The variation of parameter $\lambda$ as  a function 
        	of $\sqrt{s}$ and rapidity. The parameter $\lambda$ is 
        	parameterized with a power law of the form 
        A$\times$ $\sqrt{s}^{B}$, shown by the dashed line.
        	The parameter $k$ is extracted from the \ee 
        	data~\cite{TASSO, ALEPH}.}
\label{fig:lambda}
\end{figure}%

 The Weibull parameters $k$ and $\lambda$ are studied as a 
 function of $\sqrt{s}$ and rapidity, are shown in Figure~\ref{fig:k}
  and  Figure~\ref{fig:lambda}, respectively. 
The parameters $k$ and $\lambda$ are parameterized as a function of $\sqrt{s}$ with power law of the form A$\times \,\sqrt{s}^{\,B}$. The parameterized 
values are given in Table~\ref{t3} and Table~\ref{t4}.
\begin{table}
\caption{The values of parameters obtained for $k$ as a function of $\sqrt{s}$.}
\label{t3}
\begin{tabular}{ccc}
\hline
\hline
Parameters & $|y| < $ 0.5 & $|y| < $ 2.0 \\
\hline
      A    & 1.46 $\pm$ 0.176\,  & 2.66  $\pm$ 0.187 \\ 
      B    & -0.067 $\pm$ 0.034\,  & -0.067 $\pm$ 0.0186 \\ 
 
\hline
\hline
\end{tabular}
\end{table}

\begin{table}
\caption{The values of parameters obtained for $\lambda$ as a function of $\sqrt{s}$.}
\label{t4}
\begin{tabular}{ccc}
\hline
\hline
Parameters & $|y| < $ 0.5 & $|y| < $ 2.0\\
\hline
      A    & 1.57 $\pm$ 0.132\,  & 3.504  $\pm$ 0.292 \\ 
      B    & 0.153 $\pm$ 0.025\,  & 0.321 $\pm$ 0.0200 \\ 
 
\hline
\hline
\end{tabular}
\end{table}

  It is observed that as a function of 
 $\sqrt{s}$, the value of $\lambda$ shows a slight increase within 
 uncertainties for $|y| <$ 0.5, whereas for $|y| < $ 2.0, $\lambda$ 
 increases significantly with increasing $\sqrt{s}$. 
 As the parameter $\lambda$ is  associated with mean multiplicity~\cite{weibull1}, it
 is straightforward to see its increase with collision energies. 
   In this particular result, it can be seen
 that the value of $\lambda$ is $\sim$ 4 - 5  times higher for $|y| <$ 2.0 
 in comparison to $|y| <$ 0.5. Since the increase of $\lambda$ is not very significant 
 for $|y| <$ 0.5, one can attribute the increase for $|y| <$ 2.0 to the contribution 
 coming from ``soft" processes in forward rapidity region. 
	The value of $k$  do not vary significantly with the center of mass energy 
	for the same rapidity intervals, indicating that 
	the dynamics associated with the fragmentation process in \ee collisions 
	is very similar for the given range of energies. 
	However, the value of $k$ is higher for larger rapidity interval 
	and similar behavior was observed in hadronic collisions~\cite{weibull1}. 
	This can be related to probing a more ``soft" region where the 
	produced partons merge with very soft gluons to form hadrons,  
	usually the large mass resonances which eventually decay.

It has been observed that the average multi-hadronic final states  in
different interacting systems shows a dependence on $\sqrt{s}$. 
The observed dependence disappears if the `effective energy' 
is used to charactrize the interacting system rather than the center
of mass energy~\cite{effective_energy, effective_energy1}. Taking into account the effective energy scenario,
 the energy available for particle production
in \pp collisions is the energy of the single interacting quark pair.  As a result,
about one-third of the  entire nucleon energy is only available for particle
production in such collisions.  However, in \ee collisions, the annihilation
process utilizes the entire available collision energy for the
production of final state particles. 

Thus, one expects to observe a good agreement on  charged particle
multiplicity distributions between \ee collisions and \pp collisions
when the center of mass energy of the later  is three times of \ee collisions.
It is noteworthy to see, if the multiplicity distribution in \pp
collisions can be  obtained from the  measured multiplicity
distributions in \ee collisions and vice versa using the Weibull parameters. 

\begin{figure*}
\centering
\begin{tabular}{lr}
\begin{minipage}{0.5\textwidth}
\includegraphics[width=0.9\textwidth]{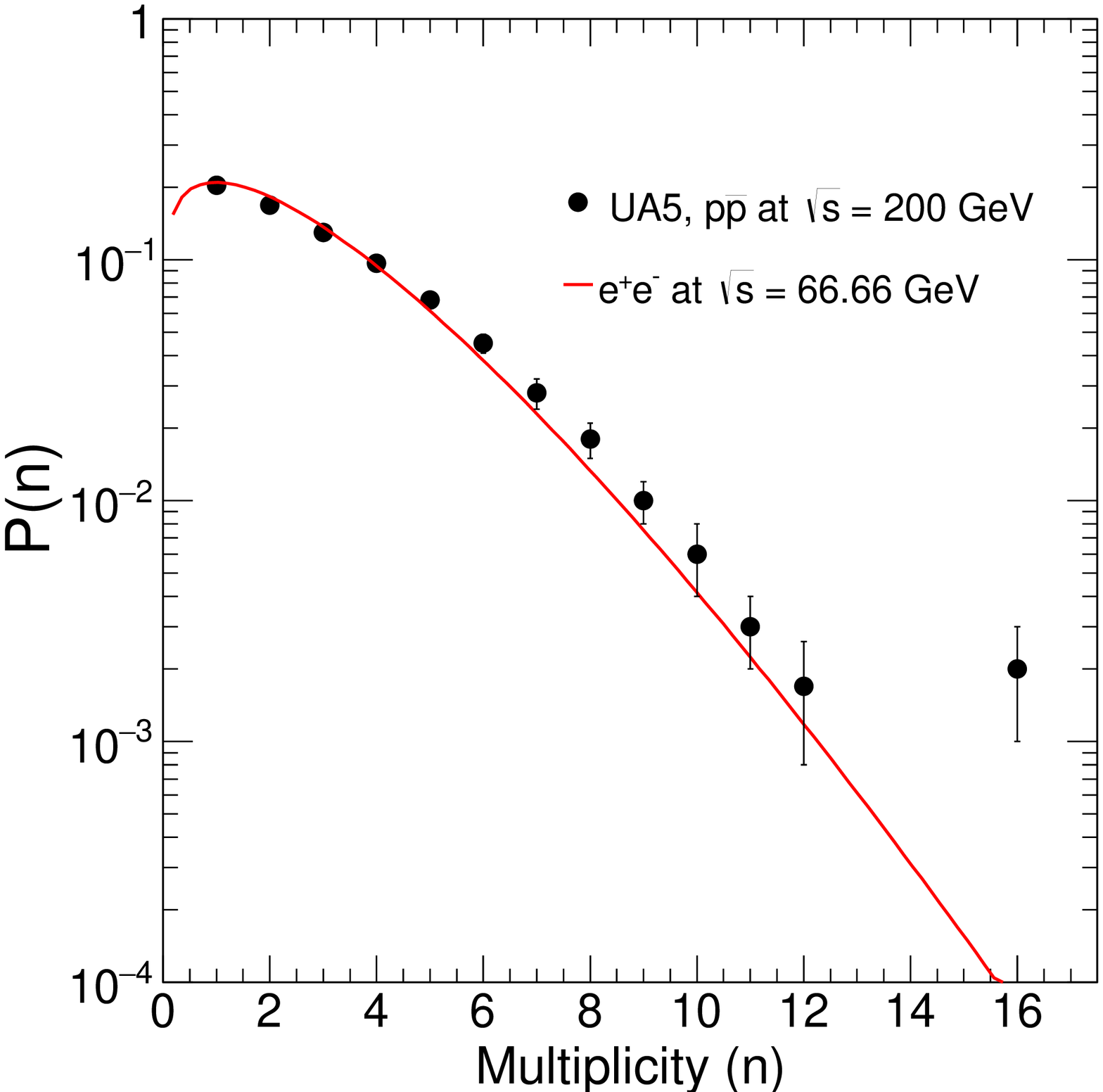}
\caption{Comparison of the multiplicity distribution
for 200 GeV $p\bar{p}$ data~\cite{ua5} (solid markers) and the multiplicity distribution obtained using Weibull parameters 
for \ee collisions (solid line) at $\sqrt{s}$ = 66.66 GeV, in 
accordance with the effective-energy approach~\cite{effective_energy}. }
\label{f5}
\end{minipage}
\begin{minipage}{0.5\textwidth}
\includegraphics[width=0.9\textwidth]{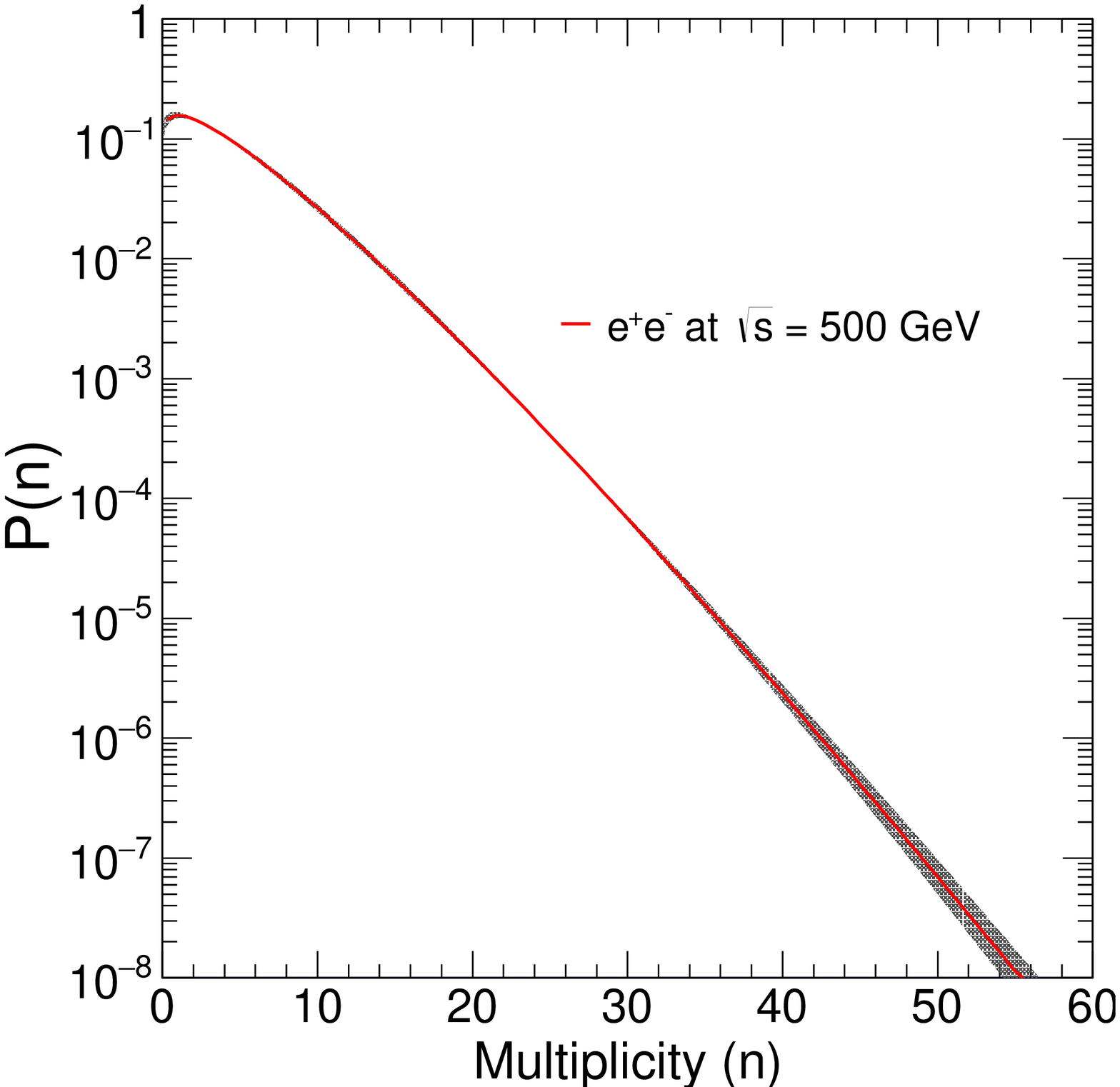}
\caption{The  charged particle multiplicity distribution
 in \ee collisions for $|{\eta}| <$ 0.5 at  $\sqrt{s}$ =  500 GeV as predicted by  Weibull parametrization. The shaded band around the solid line shows  the associated systematic errors.}
\label{f6} 
\end{minipage}
\end{tabular}
 \end{figure*}

The parameters $k$ and $\lambda$ for \ee collisions 
are studied as a function of $\sqrt{s}$. In order to verify  
the effective energy scenario, the data for $p\bar{p}$ collisions in
200 GeV is used from UA5 collaboration~\cite{ua5}. 
The values of $k$ and $\lambda$ for \ee collisions are interpolated
from Figure~\ref{fig:k} and Figure~\ref{fig:lambda} respectively for  the center of mass energy 66.66 GeV (which is one third of 200
GeV). 
Figure~\ref{f5} compares the multiplicity distribution in \ee
collisons at 66.66 GeV using Weibull parametrization with the measured
multiplicity distribution in $p\bar{p}$ collisons at 200 GeV.
One can observe an excellent agreement between the two collisions
systems favoring the  effective energy model of particle production.
This remarkable agreement  can be used  to predict the mulitplicity 
distribution for  \ee collisions at 500 GeV~\cite{ILC} as it should be similar to
the multiplicity distribution of $p$+$p$ collisions at 1500 GeV.
The multiplicity distribution for $p$+$p$ collisions  can be
obtained by interpolating  the Weibull parameters at 1500 GeV from Ref.~\cite{weibull1}.  
The $k$ and the $\lambda$ values  obtained for 
$\sqrt{s}$ = 1500 GeV are 1.17 $\pm$ 0.02 and 
4.84 $\pm$ 0.17 respectively.  The resultant
distribution is shown in Figure~\ref{f6}.

\section{Conclusion}
\label{conclusion}
The Weibull distribution provides an excellent description of the
multiplicity distributions in \ee collisions at  broad range of
energies for two extreme rapidity intervals.
The parameters of the distribution were studied as a function 
of the collision energies for two different rapidity intervals. The
$\lambda$  parameter shows a slight increase with collision  energy
while the $k$ parameter do not vary significantly with
energy. This study suggests that most of the particles  are produced
via the soft processes in forward rapidity region. Furthermore, the
effective energy model was also verified and was used to predict the
multiplicity distributions in \ee collsions at ILC energies. 
Thus, the wide applicability of the Weibull model as an effective model 
to describe the particle production in different collision systems has
been demonstrated.

\section{Acknowledgments}
We are grateful to Prof. Edward Sarkisyan-Grinbaum for fruitful comments and  suggestions.

\end{document}